\documentstyle[twoside,fleqn,espcrc2,epsfig]{article}

\title{
Testing the Leutwyler-Smilga prediction regarding the global topological charge
distribution on the lattice\thanks{Talk given at
QCD'00, Montpellier, 6-12 July 2000.}}

\author{S. D\"urr
\address{Paul Scherrer Institut, Theory Group, 5232 Villigen PSI, Switzerland}
\thanks{E-mail: {\tt stephan.duerr@psi.ch}}}

\begin{document}

\begin{abstract}
I give a sketch of my recent attempt to test the prediction by Leutwyler and
Smilga according to which, for QCD in a finite box with $N_{\!f}\!\geq\!2$,
the combination $x\!=\!V\Sigma m$
indicates whether the net topological charge of the gauge background proves
relevant ($x\ll1$) or irrelevant ($x\gg1$) for physical observables.
\vspace*{-3mm}
\end{abstract}

\hyphenation{ob-vious propa-ga-tors resul-ting vacu-um simu-la-tion pheno-menon
pheno-mena renormali-za-tion con-figura-tion topolo-gi-cal topolo-gi-cal-ly
inter-me-diate pheno-meno-logi-cal typi-cal}

\maketitle

\def\J#1#2#3#4{{#1} {\bf #2} (#4), #3}
\def\NPB{{\sl Nucl.\ Phys.} {\bf B}}
\def\NPP{\sl Nucl.\ Phys.\ (Proc.\ Suppl.)}
\def\PLB{{\sl Phys.\ Lett.} {\bf B}}
\def\PRD{{\sl Phys.\ Rev.} {\bf D}}
\def\PRL{\sl Phys.\ Rev.\ Lett.}
\def\CMP{\sl Comm.\ Math.\ Phys.}
\def\AOP{\sl Ann.\ Phys.}
\def\JPA{{\sl J.\ Phys.} {\bf A}}
\def\JPQ{\sl J.\ Physique (France)}
\def\HPA{\sl Helv.\ Phys.\ Acta}
\def\ZPC{{\sl Z.\ Phys.} {\bf C}}
\def\JHEP{\sl J.\ High Energy Phys.}

\newcommand{\pad}{\partial}
\newcommand{\pas}{\partial\!\!\!/}
\newcommand{\Dsl}{D\!\!\!\!/\,}
\newcommand{\Psl}{P\!\!\!\!/\;\!}
\newcommand{\Nf}{N_{\!f}}
\newcommand{\hqu}{\hbar}
\newcommand{\ovr}{\over} 
\newcommand{\hal}{{1\ovr2}}
\newcommand{\til}{\tilde}
\newcommand{\pri}{^\prime}
\renewcommand{\dag}{^\dagger}
\newcommand{\<}{\langle}
\renewcommand{\>}{\rangle}
\newcommand{\gaf}{\gamma_5}
\newcommand{\lap}{\triangle}
\newcommand{\trc}{{\rm tr}}
\newcommand{\al}{\alpha}
\newcommand{\be}{\beta}
\newcommand{\ga}{\gamma}
\newcommand{\de}{\delta}
\newcommand{\ep}{\epsilon}
\newcommand{\ve}{\varepsilon}
\newcommand{\ze}{\zeta}
\newcommand{\et}{\eta}
\newcommand{\th}{\theta}
\newcommand{\vt}{\vartheta}
\newcommand{\io}{\iota}
\newcommand{\ka}{\kappa}
\newcommand{\la}{\lambda}
\newcommand{\rh}{\rho}
\newcommand{\vr}{\varrho}
\newcommand{\si}{\sigma}
\newcommand{\ta}{\tau}
\newcommand{\ph}{\phi}
\newcommand{\vp}{\varphi}
\newcommand{\ch}{\chi}
\newcommand{\ps}{\psi}
\newcommand{\om}{\omega}
\newcommand{\psb}{\overline{\psi}}
\newcommand{\etb}{\overline{\eta}}
\newcommand{\psd}{\psi^{\dagger}}
\newcommand{\chd}{\chi^{\dagger}}
\newcommand{\etd}{\eta^{\dagger}}
\newcommand{\etp}{\eta^{\prime}}
\newcommand{\rch}{{\rm ch}}
\newcommand{\rsh}{{\rm sh}}
\newcommand{\lab}{\overline{\la}}
\renewcommand{\i}{{\rm i}}
\newcommand{\qal}{q_\al}
\newcommand{\qbe}{q_\be}
\newcommand{\qmu}{q_\mu}
\newcommand{\qnu}{q_\nu}
\newcommand{\gmunu}{g_{\mu\nu}}
\newcommand{\Afin}{A_{{\rm fin}}}
\newcommand{\Bbar}{\overline B}
\newcommand{\Jbar}{\overline J}
\newcommand{\Bmu}{B_\mu}
\newcommand{\Bnu}{B_\nu}
\newcommand{\Bmunu}{B_{\mu\nu}}
\newcommand{\Talmu}{T_{\al\mu}}
\newcommand{\Talnu}{T_{\al\nu}}
\newcommand{\Tmunu}{T_{\mu\nu}}
\newcommand{\Cmu}{C_\mu}
\newcommand{\Cnu}{C_\nu}
\newcommand{\Cmunu}{C_{\mu\nu}}
\newcommand{\Umunu}{U_{\mu\nu}}
\newcommand{\Fsq}{F_0^2}
\newcommand{\psq}{p^2}
\newcommand{\qsq}{q^2}
\newcommand{\msq}{m^2}
\newcommand{\mo}{m_1^2}
\newcommand{\mt}{m_2^2}
\newcommand{\Mo}{M_1^2}
\newcommand{\Mt}{M_2^2}
\newcommand{\Met}{M_\et}
\newcommand{\Mka}{M_K}
\newcommand{\Mpi}{M_\pi}
\newcommand{\Mesq}{M_\et^2}
\newcommand{\Mksq}{M_K^2}
\newcommand{\Mpsq}{M_\pi^2}
\newcommand{\mesq}{M_\et^2}
\newcommand{\mksq}{M_K^2}
\newcommand{\mpsq}{M_\pi^2}
\newcommand{\beq}{\begin{equation}}
\newcommand{\eeq}{\end{equation}}
\newcommand{\bdm}{\begin{displaymath}}
\newcommand{\edm}{\end{displaymath}}
\newcommand{\bea}{\begin{eqnarray}}
\newcommand{\eea}{\end{eqnarray}}

\setcounter{footnote}{0}


\begin{figure}[h]
\vspace*{-6mm}
\includegraphics{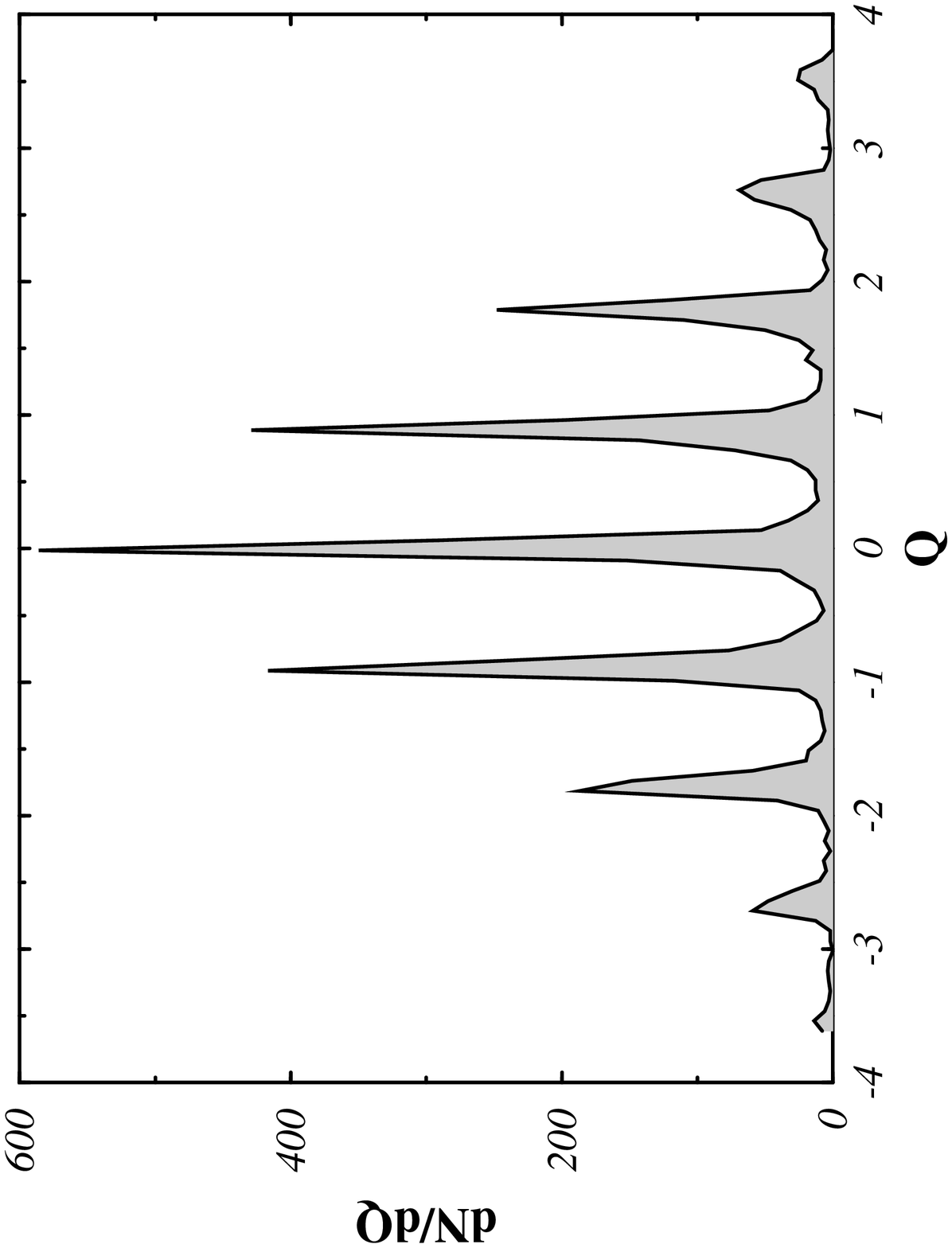}
\vspace*{45mm}
\includegraphics{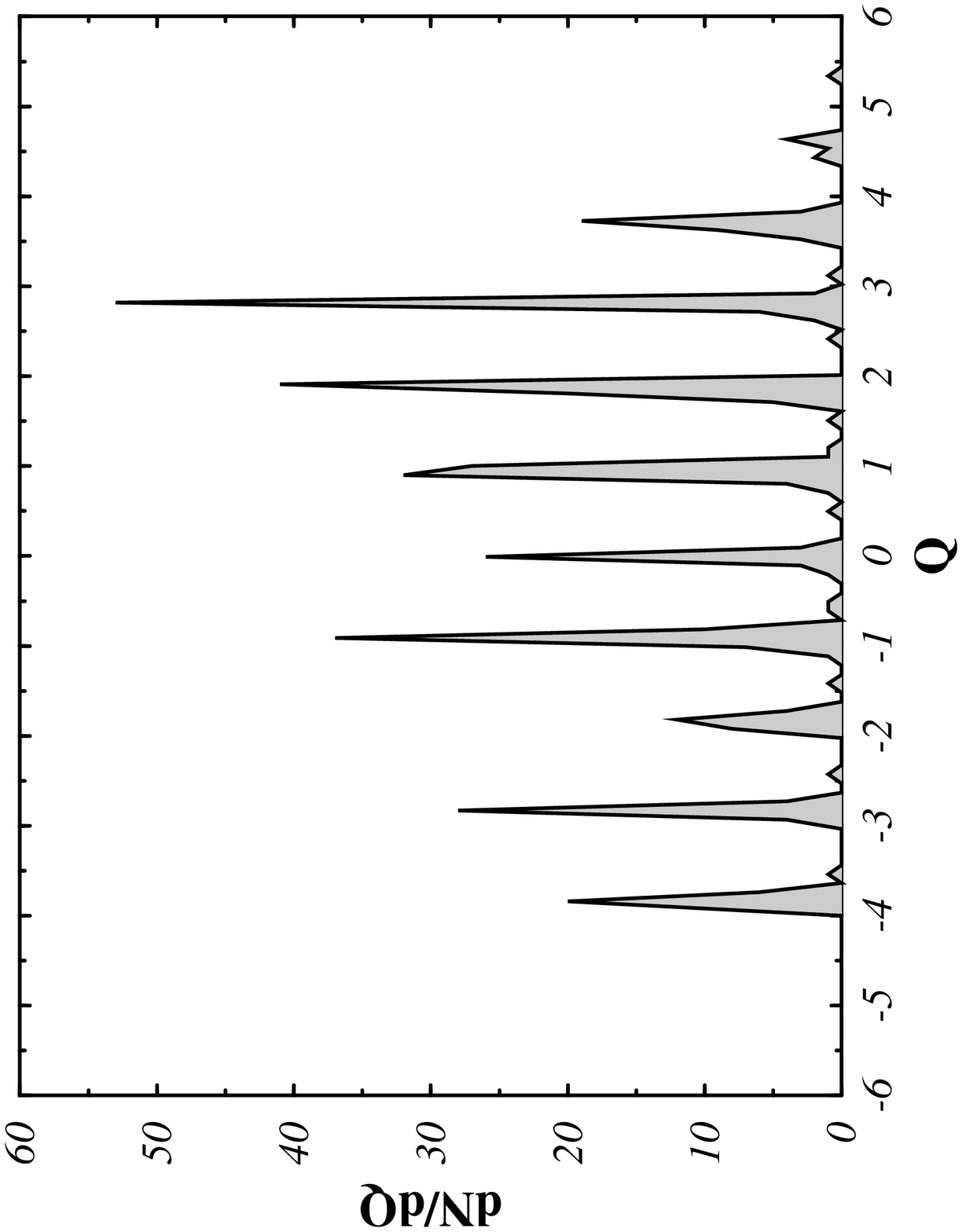}
\vspace*{45mm}
\caption{Distribution of the naive topolo\-gical charge $\nu_{\rm nai}$ (see
footnote 1) in an
ensemble of $SU(3)$-configurations generated with $S_{\tiny\rm Wilson}$ at
$\beta\!=\!6.1$ (top) and in an ensemble of (full) QCD configurations (bottom).
Figures taken from \cite{AllesICHEP96}.}
\vspace*{-11mm}
\end{figure}

\section{INTRODUCTION}

QCD in a finite box ---~or, more generally, the finite-volume version of any
field theory which breaks a continuous global symmetry spontaneously~---
provides a rich and fascinating subject: One expects to see both symmetry
restoration phenomena and the onset of spontaneous symmetry breaking (e.g.\
of the global $SU(N_{\!f})_A$ symmetry in QCD), if the 
box volume is taken sufficiently ``small'' or ``large'', respectively.
An obvious question is what sets the scale, i.e.\ by which
standards does the box have to be small or large to
trigger one or the other type of phenomena~?
The naive guess is of course the lightest particle
at hand, i.e.\ the mass $M_\pi$ of the Goldstone boson produced in the infinite
volume limit.
From this one would expect symmetry restoration phenomena to be pronounced
for $M_\pi L\!\ll\!1$ and SSB to become manifest for $M_\pi L\!\gg\!1$.

Another place where finite and infinite volume aspects of the theory meet
is the question whether the total topological charge of the gauge field
\beq
\nu={g^2\ovr 32\pi^2}\int G\til G\ dx
\label{nudef}
\eeq
(which is a finite volume concept) proves relevant in physical observables
(after extrapolating to infinite volume) or not.
From a lattice perspective this question deserves attention, because of a
technical subtlety with numerical QCD studies:
In quenched simulations, excellent ergodicity of the sample w.r.t.\ the
topological charge is easy to achieve (see top part in Fig.\ 1), i.e.\ the
overall distribution%
\footnote{On the lattice a naive implementation of (\ref{nudef}) ---~the result
being called $\nu_{\rm nai}$~--- does not yield an integer, but a distribution
which gets closer to a set of delta peaks near integer numbers only in the
continuum limit $a\!\to\!0$. In the following, our interest is in the overall
distribution, i.e.\ in the embedding curve of the distribution shown in
Fig.\ 1}
is typically symmetric and centered about zero.
In a dynamical run (one that includes the fermion determinant) the resulting
distribution may look like the one shown in the lower part of Fig.\ 1 --- the
simulation seems far from having achieved a good sampling of the topological
charge.
There are two reasons responsible for this:
First, standard full QCD algorithms have a tendency to ``get stuck'' in a
particular topological sector, if the quark mass is taken sufficiently light.
This holds true both with the staggered~\cite{TopErgodicityHMCstag} and with
a Wilson type formulation~\cite{TopErgodicityHMCwils} of the (dynamical)
fermions.
Second, because of the tremendous increase in costs (in terms of CPU time) per
configuration, the total sample weight is typically smaller than in the
quenched case.
Hence the physical question is whether the full QCD sample shown in the lower
part of Fig.\ 1 results in unbiased measurements of $M_\pi, F_\pi$ etc.\ or
whether the supposed overrepresentation of the sector with $\nu\!=\!3$ (see
Fig.\ 1) is likely to affect the extraction of physical quantities.


\section{RESULTS OF THE ANALYSIS BY LEUTWYLER AND SMILGA}

A key point in the analysis by Leutwyler and Smilga \cite{LeutwylerSmilga}
is that the topics raised in the introduction are closely interwoven and that
the discussion involves yet another parameter%
\footnote{Here and below $V$ denotes the four-volume of the box,
$\Sigma\!=\!-\!\lim_{m\to 0} \lim_{V\to\infty} \langle\bar\psi\psi\rangle$
(this order) is the chiral condensate in the chiral limit and $m$ is the
(degenerate) {\em sea\/}-quark mass; both $\Sigma$ and $m$ are scheme- and
scale-dependent, but the combination is an RG-invariant.}
\beq
x\equiv V\Sigma m
\label{LSPdef}
\eeq
which really decides whether the box is ``small'' or ``large''.
This new parameter is, in principle, independent of the naive parameter
$M_\pi L$:
In the regimes $x\!\gg\!1$ and $x\!\simeq\!1$ with pronounced or mild SSB the
box may be large ($M_\pi L\!\gg\!1$), intermediate ($M_\pi L\!\simeq\!1$) or
small ($M_\pi L\!\ll\!1$) w.r.t.\ the conventional classification, whereas in
the symmetry restoration regime ($x\!\ll\!1$) such a distinction does not make
sense anyways, because there the pion is not a useful degree of freedom.
Note that in a lattice context the Leutwyler-Smilga (LS) classification refers
to the mass of the sea-quarks (i.e.\ the dynamical quarks which influence the
functional weight of a gauge configuration through the determinant), whereas
the conventional one refers to the mass of the current-quarks (i.e.\ those from
which the observable is built).

The main result of the analysis by Leutwyler and Smilga \cite{LeutwylerSmilga}
is that there are two regimes of quark masses and box volumes for which the
path-integral may be evaluated analytically and statements regarding the
relevance of the topological charge $\nu$ can be formulated.
They find that in the symmetry restoration regime ($x\!\ll\!1$) where the
fundamental description in terms of quarks and gluons is appropriate the
partition function $Z$ is completely dominated by the contribution $Z_0$ from
the topologically trivial sector.
In the opposite regime ($x\!\gg\!1$) where SSB becomes manifest (though
formally the box volume is still finite) an effective description in terms of
Goldstone degrees of freedom proves useful, but here two subcases must be
treated separately:
For $M_\pi L\!\gg\!1$ (i.e.\ if the pion stays inside the box) standard chiral
perturbation theory may be used with the obvious replacement
${1\over p^2+M_\pi^2}\!\to\!\sum{1\over p^2+M_\pi^2}$, i.e.\ the propagators
should account for the ``mirror copies''.
Since this perturbative evaluation assumes a single vacuum (sc.\ the one with
$\nu\!=\!0$) no statement regarding the $\nu$-dependence can be made for this
subcase.
For $M_\pi L\!\ll\!1$ (i.e.\ if the pion overlaps the box) the zero-modes
provide the dominant contribution to the path-integral in the effective
description \cite{GLthermodynamics}, and Leutwyler and Smilga end up finding
that the partition function is approximately independent of $\nu$ or, more
precisely, that the distribution is \cite{LeutwylerSmilga}
($x\!\gg\!1, M_\pi L\!\ll\!1$)
\beq
Z_\nu \sim e^{-{\nu^2\ovr2\<\nu^2\>}}
\quad{\rm with}\quad
\<\nu^2\>\!=\!{V\Sigma m\ovr N_{\!f}}
\;.
\label{LSgaussian}
\eeq

This is nice, but from a lattice perspective the statement about the
$\nu$-independence of the partition function in the large $x$ regime is exactly
in the wrong subcase --- in numerical simulations the pion would always fit
into the box.
Hence, besides testing the LS-prediction for the small $x$ regime, an obvious
goal was to study the $\nu$-dependence of both the partition function and
selected observables in the regimes with mild ($x\!\simeq\!1$) or pronounced
($x\!\gg\!1$) SSB after reversing the ``overlap-condition'' into
$M_\pi L\!\gg\!1$ \cite{Durr}.


\begin{figure*}[t]
\hspace*{1mm}
\epsfig{file=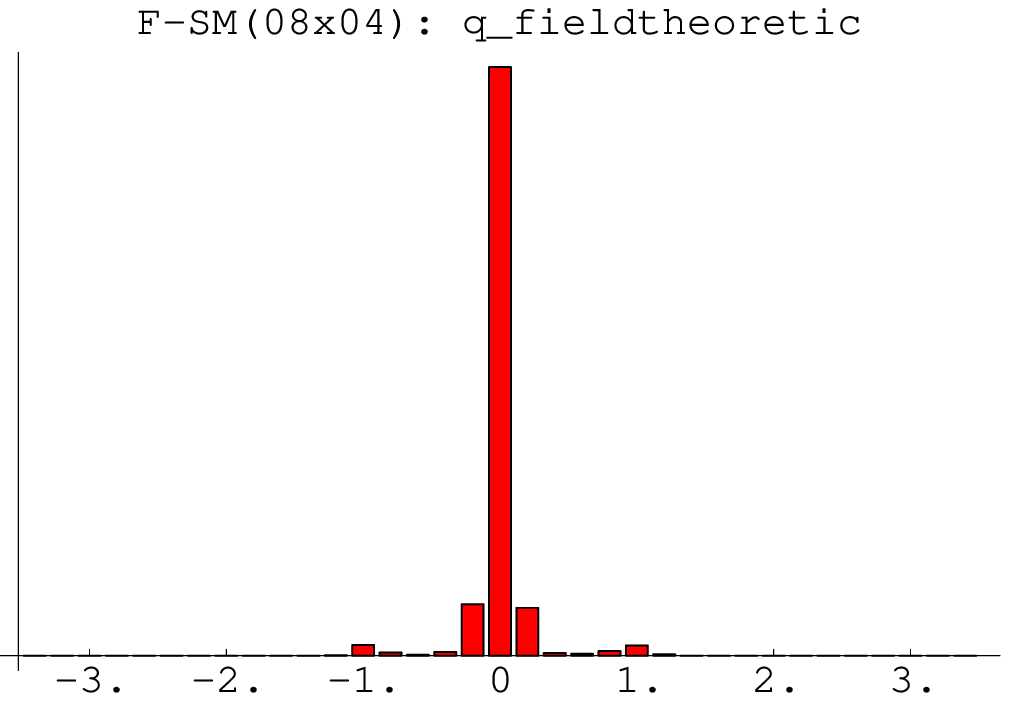,width=5.0cm}
\hspace*{1mm}
\epsfig{file=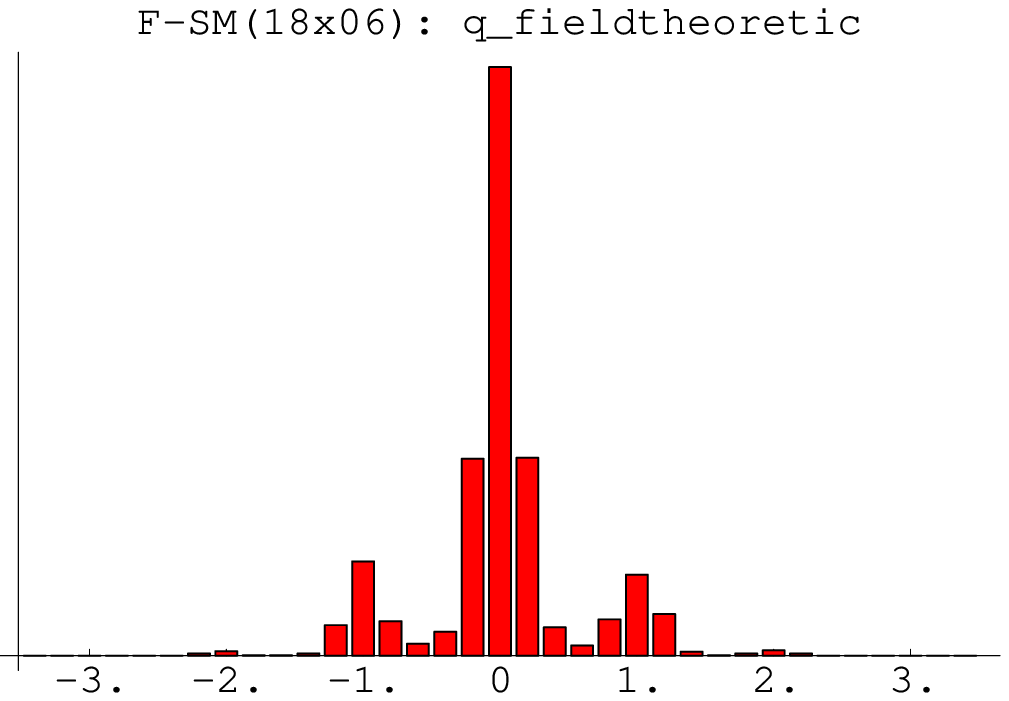,width=5.0cm}
\hspace*{1mm}
\epsfig{file=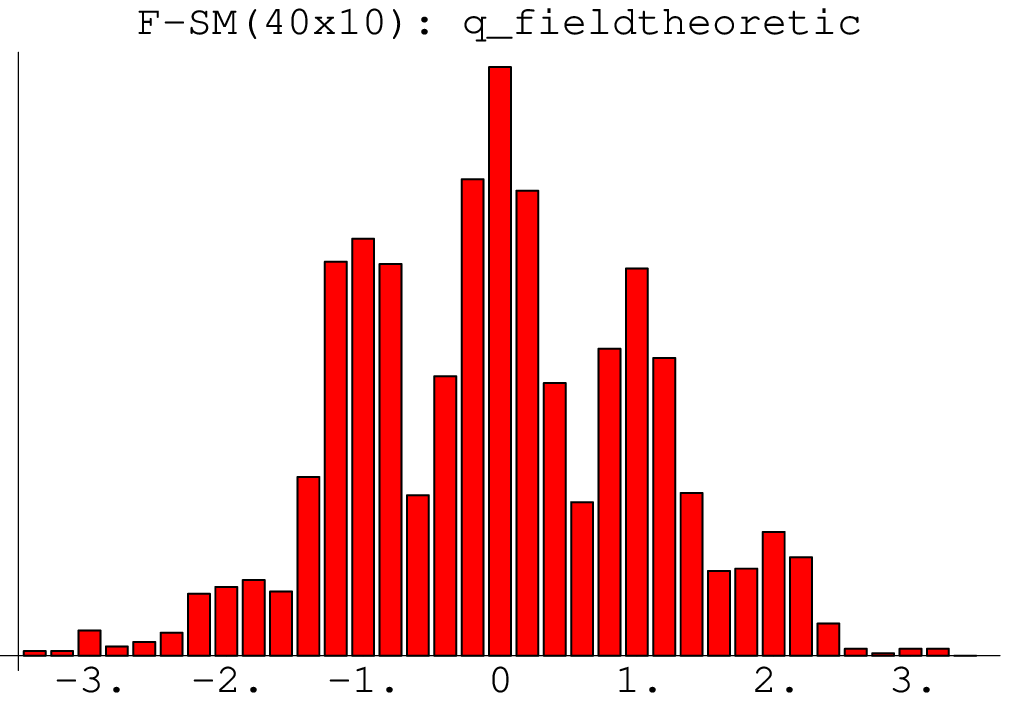,width=5.0cm}
\\
\hspace*{1mm}
\epsfig{file=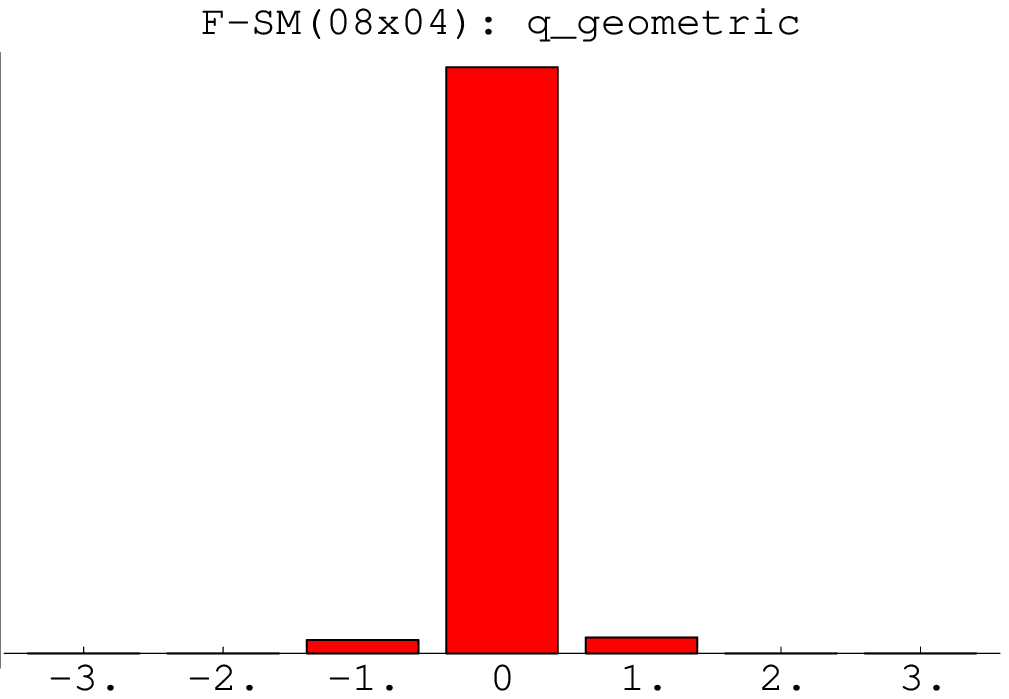,width=5.0cm}
\hspace*{1mm}
\epsfig{file=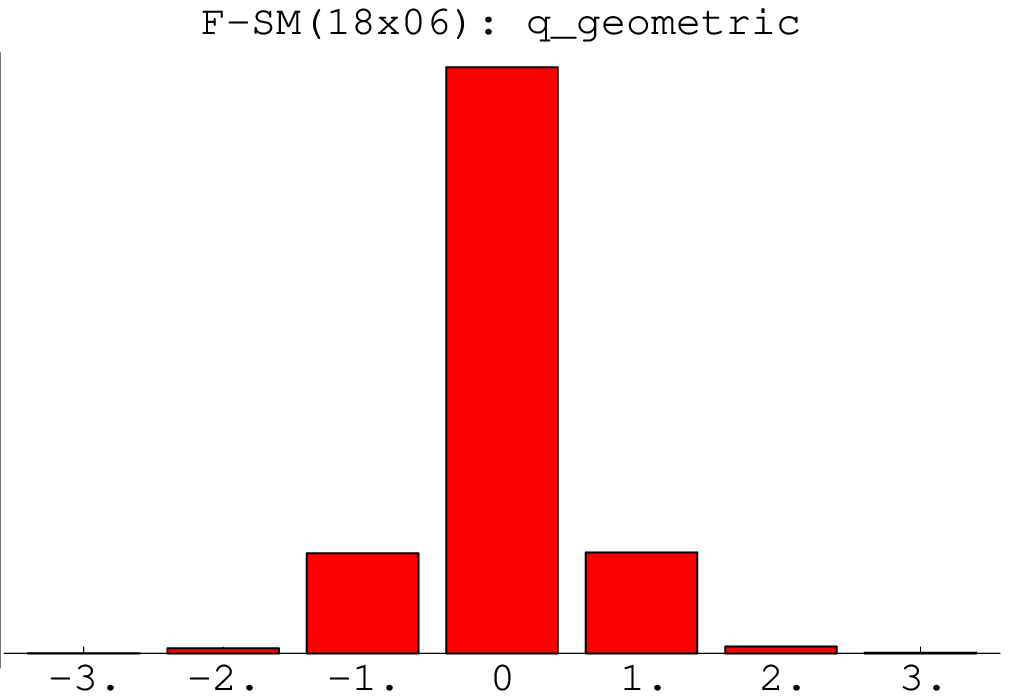,width=5.0cm}
\hspace*{1mm}
\epsfig{file=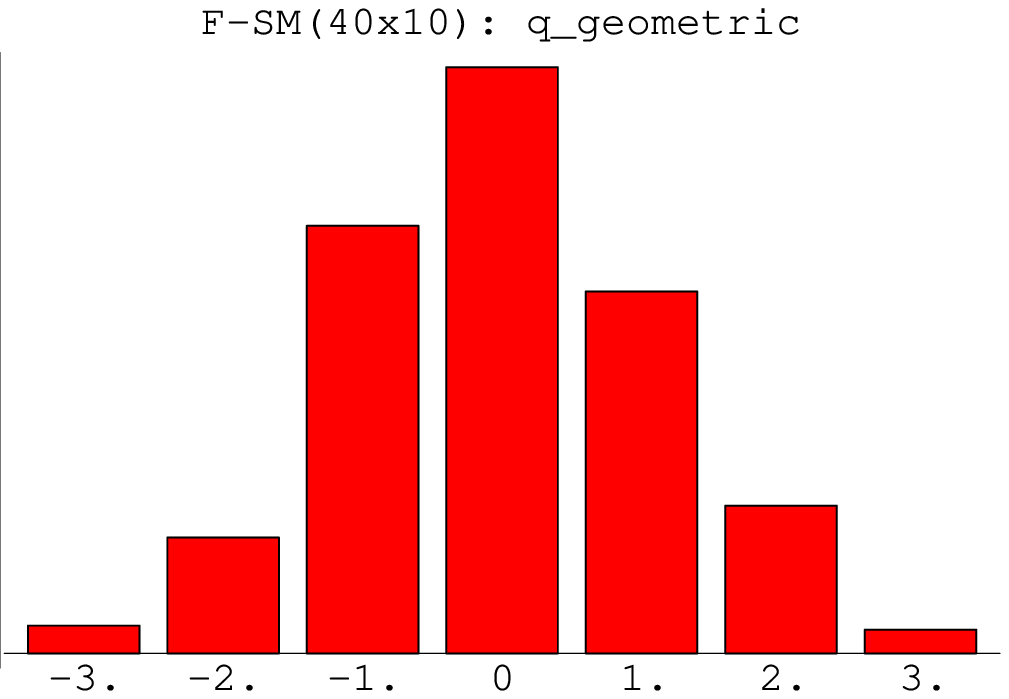,width=5.0cm}
\vspace*{-9mm}
\caption{
Distribution of $\nu_{\rm fth}$ (top) and $\nu_{\rm geo}$ (bottom) in
the small (8x4 lattice), intermediate (18x6) and large (40x10)
Leutwyler-Smilga regimes, respectively
(QED(2) data for $\beta\!=\!3.4, m\!=\!0.09, N_{\!f}\!=\!2$).}
\end{figure*}

\section{CHECKING THE NET TOPOLOGICAL CHARGE DISTRIBUTION}

In order to check the LS-prediction regarding the distribution of the net
topological charge $\nu$ in the regimes $x\!\ll\!1$ and $x\!\gg\!1$, one needs
dynamical (i.e.\ unquenched) configurations.
Given the alternative to leave the subject to the phenomenological lattice
groups which are able to generate full QCD ensembles or to study it in a
suitably chosen model, I have opted for the latter possibility.
For reasons to be suppressed due to limitations of space the massive
multiflavour Schwinger model (i.e.\ QED in 2 dimensions with $\Nf\!\geq\!2$) is
supposed to reproduce all qualitative aspects of the Leutwyler Smilga issue ---
on this point the reader is referred to \cite{Durr}.
My implementation uses the Wilson gauge action
$S_{\rm gauge}\!=\!\beta\sum(1\!-\cos\theta_\Box)$ and a pair of
(dynamical) staggered fermions.

The idea is to compare the three regimes $x\!\ll\!1$, $x\!\simeq\!1,
x\!\gg\!1$ to each other using three dedicated simulations:
Working at fixed $\beta\!=\!1/(ag)^2\!=\!3.4$ and fixed staggered quark mass
$m\!=0.09$ (in lattice units), the three regimes are represented by the three
volumes $V\!=\!8\!\times\!4, 18\!\times\!6, 40\!\times\!10$, respectively.
From this setup the LS-parameter takes the values $x\!\simeq\!0.33, 1.12,
4.16$, respectively, and the pion (pseudo-scalar iso-triplet) has a (common)
mass $M_\pi\!=\!0.329$ and hence a correlation length $\xi_\pi\!=\!3.04$ as to
fit into the box.

For the type of investigation I am aiming at configurations must be assigned
an index $\nu$. I have implemented both the geometric definition
$\nu_{\rm geo}\!=\!{1\over2\pi}\sum\log U_\Box$ and the field-theoretic
version $\nu_{\rm fth}\!=\!\kappa\,\nu_{\rm nai},
\nu_{\rm nai}\!=\!\sum\sin\theta_\Box$ with the renormalization
$\kappa\!\simeq\!1/(1-\langle S_{\rm gauge}\rangle/\beta V)$~\cite{LueSmiVin}.
A configuration is assigned an index only if the geometric and the
field-theoretic definition, after rounding to the nearest integer, agree.
This turned out to be the case at a 99.9\%, 98.8\%, 88.2\% level on the
small/intermediate/large lattice, respectively.
This fraction being so high means that in practice an assignment can be done
without cooling for the overwhelming majority of configurations; the
remaining ones are just not assigned an index.

As one can see from Fig.\ 2, the charge distributions found in the three runs
seem to follow a general pattern consistent with the predictions by Leutwyler
and Smilga:
For $x\!\ll\!1$ the distribution (an hence the partition function) is very much
dominated by the contribution from the topologi\-cally trivial sector, whereas
for $x\!\gg\!1$ the distribution gets broad and seems compatible with the
gaussian form (\ref{LSgaussian}) with $\<\nu^2\>\!=\!x/2$.


\begin{figure*}[t]
\hspace*{1mm}
\epsfig{file=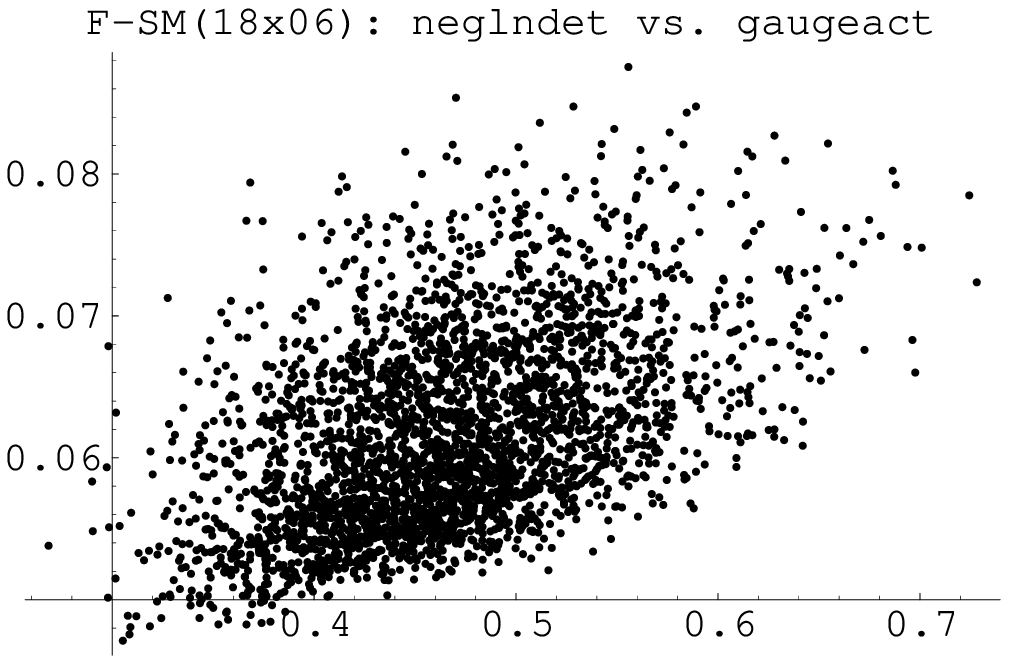,width=4.8cm,height=3.2cm}
\hspace*{1mm}
\epsfig{file=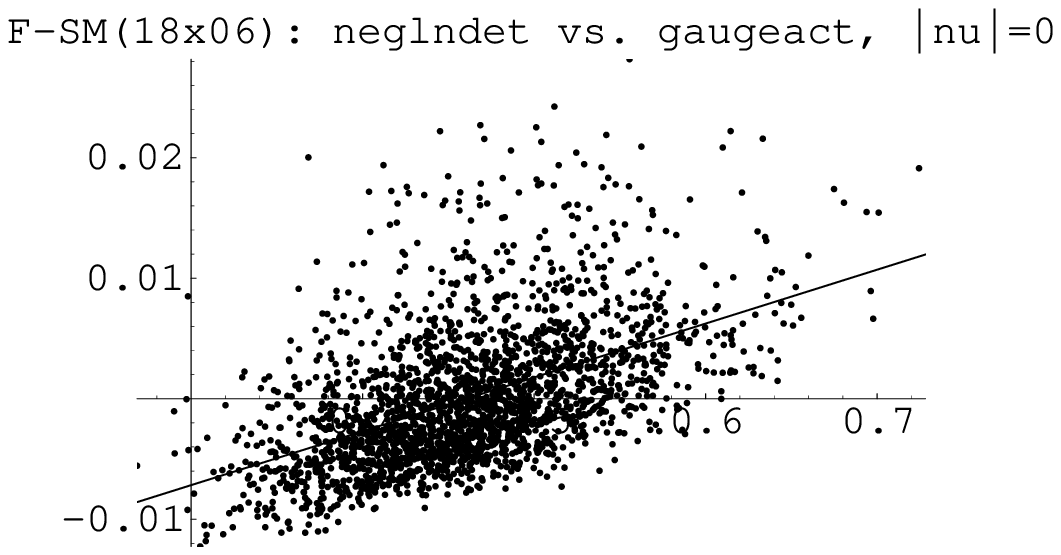,width=5.1cm,height=3.2cm}
\hspace*{1mm}
\epsfig{file=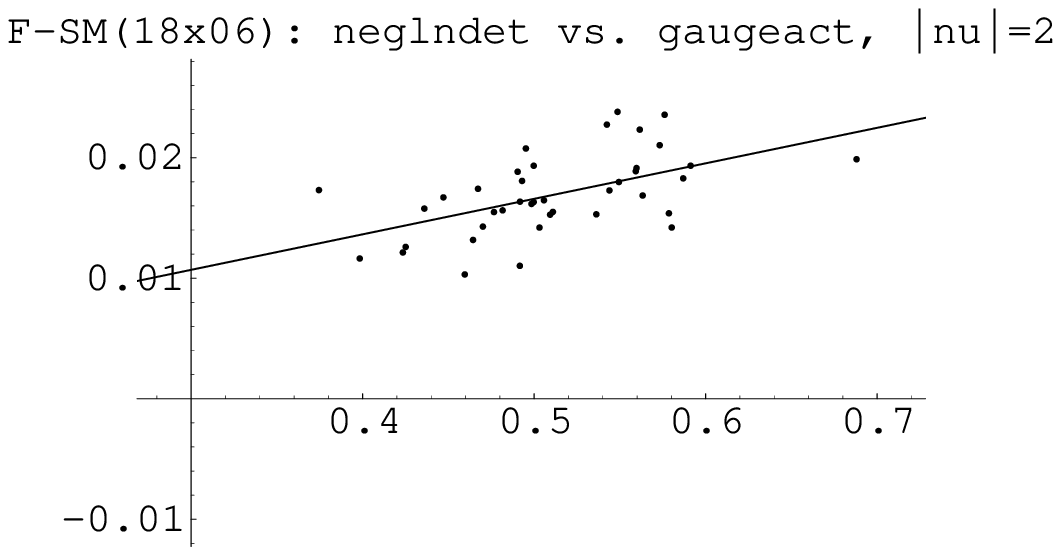,width=5.1cm,height=3.2cm}
\vspace*{-8.3mm}
\caption{
Scatter plots of $S_{\rm fermion}\!=\!-\log(\det(\Dsl\!+\!m))$ (i.e.\ per
continuum flavour) versus $S_{\rm gauge}$
(both normalized per plaquette $\Box$) on the complete sample (left) and
for the sectors with $\nu\!=\!0$ and $\nu\!=\!\pm2$ in the simulation covering
the intermediate regime --- each dot represents a configuration.}
\end{figure*}
\begin{figure*}[t]
\hspace*{1mm}
\epsfig{file=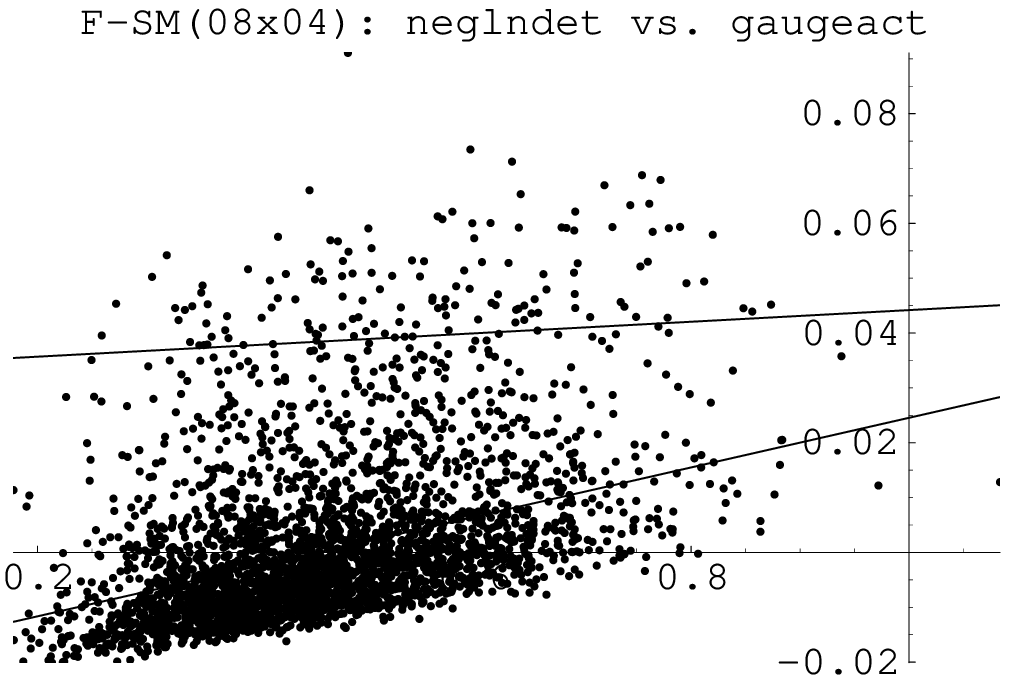,width=4.8cm,height=3.2cm}
\hspace*{1mm}
\epsfig{file=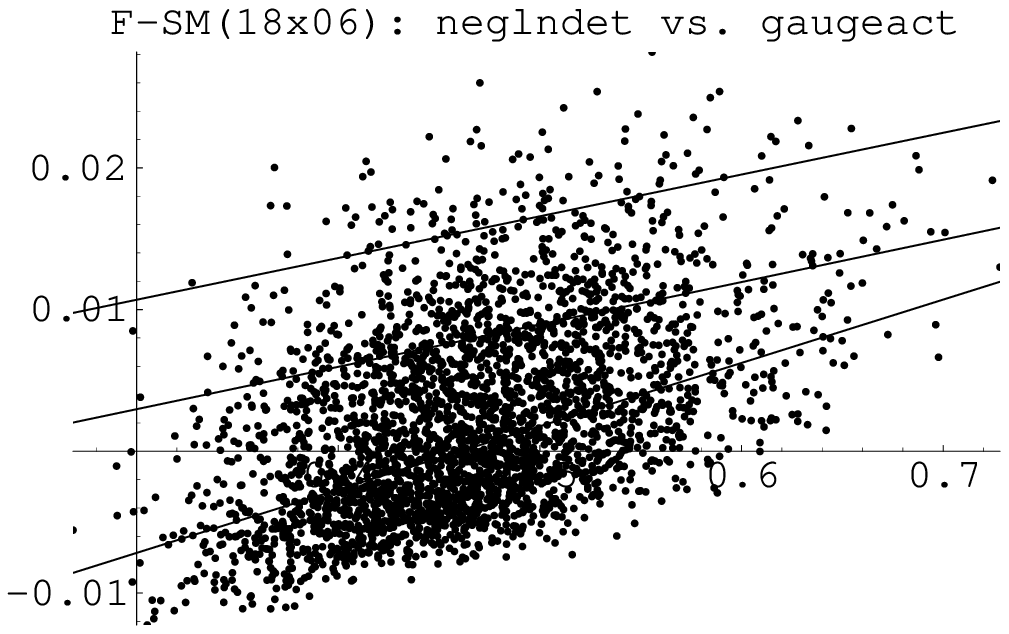,width=5.1cm,height=3.2cm}
\hspace*{1mm}
\epsfig{file=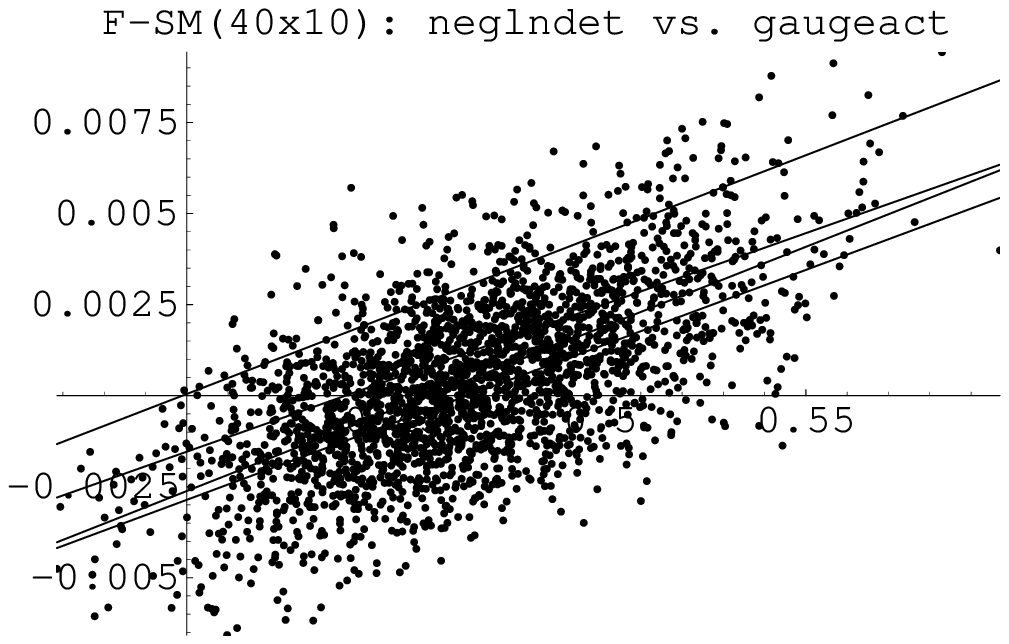,width=5.1cm,height=3.2cm}
\vspace*{-8.3mm}
\caption{
$S_{\rm fermion}$ versus $S_{\rm gauge}$ together with the best linear fits for
the reasonably populated topological sectors in the small (8x4 lattice),
intermediate (18x6) and large (40x10) Leutwyler-Smilga regimes, respectively
(QED(2) data for $\beta\!=\!3.4, m\!=\!0.09, N_{\!f}\!=\!2$).}
\end{figure*}

\section{ROLE OF THE FUNCTIONAL DETERMINANT}

Since the LS-issue is peculiar to the full (unquenched) theory, an attempt to
understand by which mechanism the three regimes differ from each other may lead
one to investigate how the functional determinant or specifically its
contribution to the total action per continuum-flavour
\beq
S_{\rm fermion}=-\log(\det(D\!\!\!\!/+\!m))+{\rm const}
\label{Sfermion}
\eeq
relates to the contribution from the gauge field and, in addition, how this
might depend on the topological charge of the background.
The idea is thus to study such a relationship sectorally, i.e.\ after the
complete sample has been separated into subsamples with a fixed topological
charge (or fixed $|\nu|$), in the spirit of Ref.\ \cite{Damgaard}.

The first plot in Fig.\ 3 (which is for $x\!\simeq\!1$) tells us that
there is a positive but weak correlation between $S_{\rm fermion}$ and
$S_{\rm gauge}$.
From this one concludes that the functional determinant acts --~roughly~--
like an effective renormalization of $\beta$ with a factor bigger than 1.
A key observation is that the correlation improves, if one separates the
sample into {\em subsamples with fixed\/} $|\nu|$, as is done in the r.h.s.\
of Fig.\ 3.
Most notably the two sectors shown ($\nu\!=\!0$ and $\nu\!=\!\pm2$) differ both
in {\em offset\/} and {\em slope\/} of the best linear fit to $S_{\rm fermion}$
versus $S_{\rm gauge}$.
This means that the functional determinant brings --~in general~-- an
{\em overall suppression\/} of higher topological sectors w.r.t.\ lower ones
and a {\em sectorally different renormalization\/} of $\beta$.

Fig.\ 4 allows one to asses the strength of these effects as $x$ varies:
In the small LS-regime both the sectoral dependence of the renormalization
factor and the offset between neighboring sectors are huge, i.e.\ for
$x\!\to\!0$ the functional determinant acts as a constraint to the
topologically trivial sector.
For $x\!\simeq\!1$ both offset and sectoral dependence of the slope are
very much reduced.
In the large LS-regime there is only a minor overall suppression of higher
topological sectors w.r.t\ lower ones, but the renormalization of $\beta$
(and hence, in 4 dimensions, the lattice spacing in physical units) seems to
be {\em uniform\/} for all sectors.


\begin{figure*}[t]
\epsfig{file=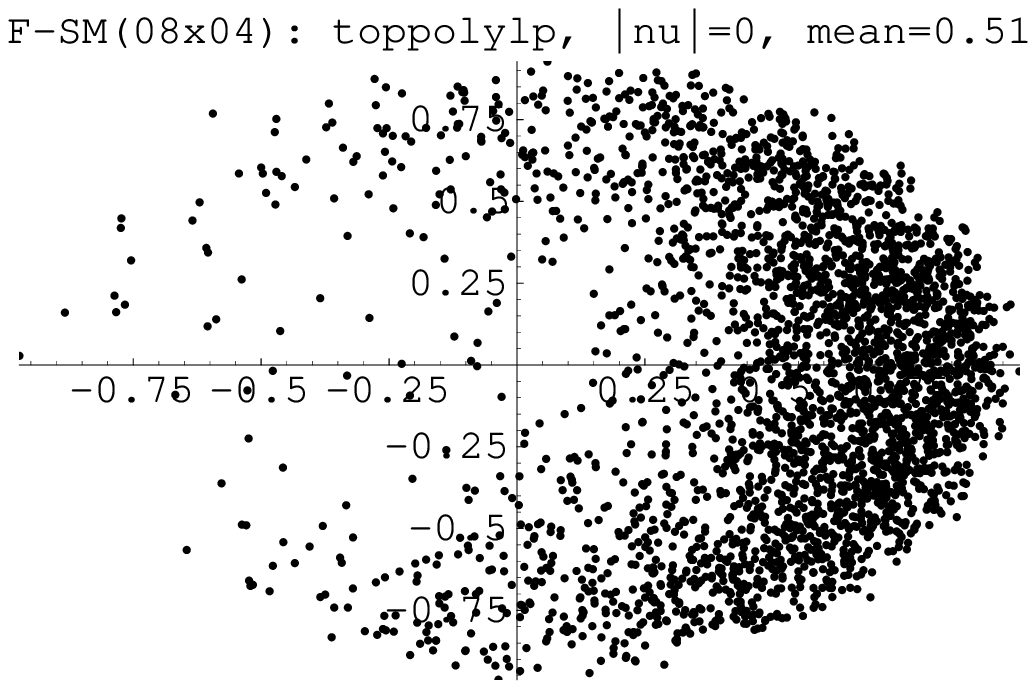,width=5.0cm}
\hspace*{1mm}
\epsfig{file=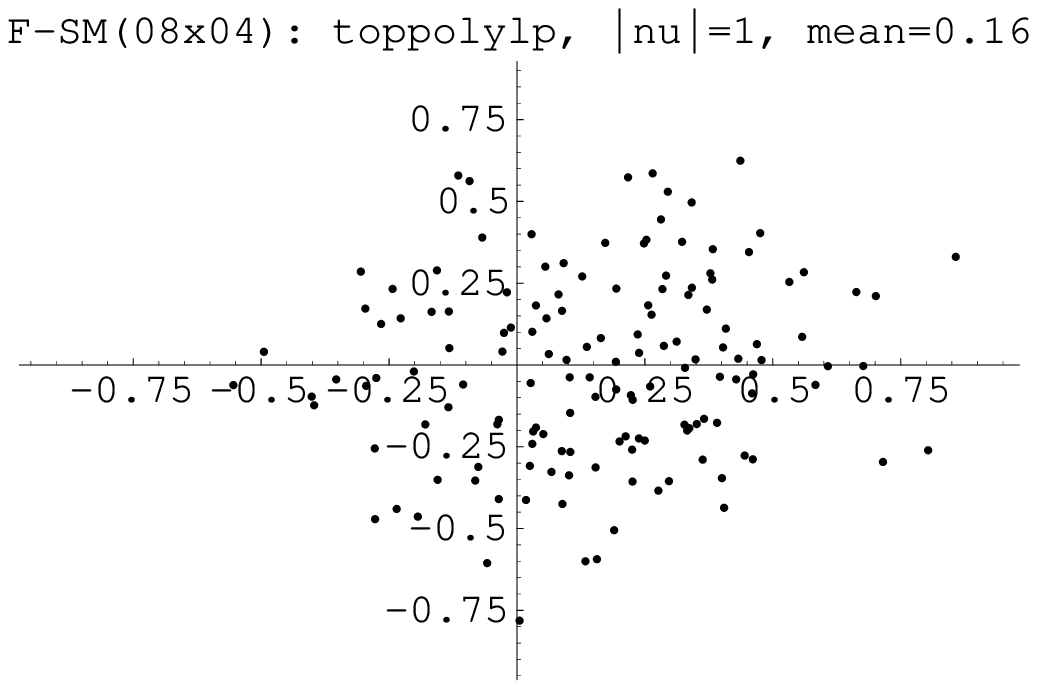,width=5.0cm}
\hspace*{1mm}
\epsfig{file=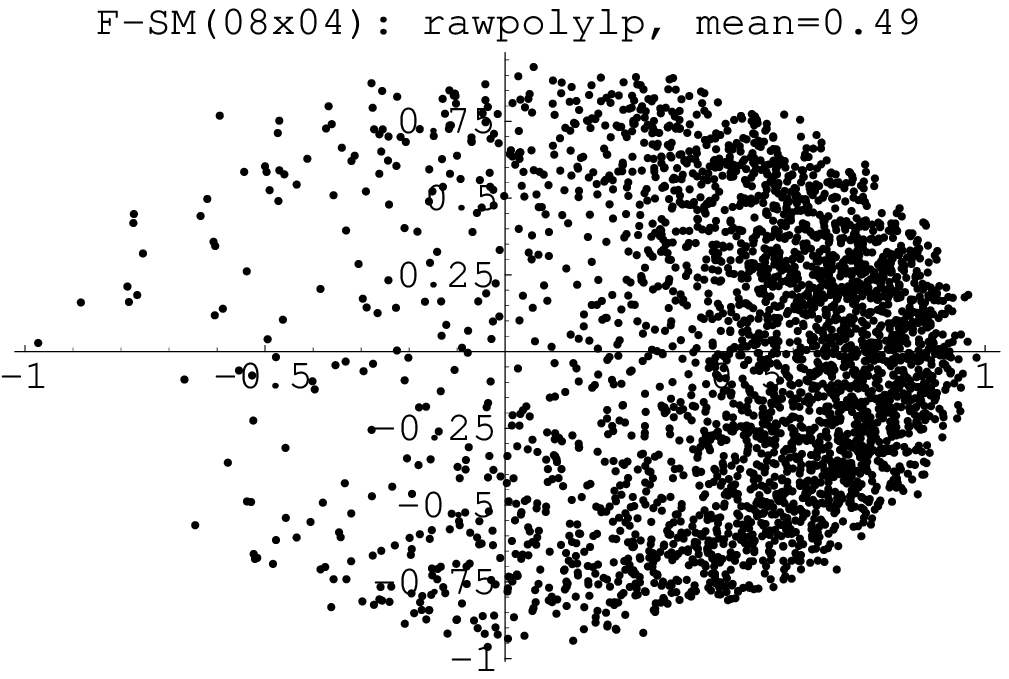,width=5.0cm}
\\
\epsfig{file=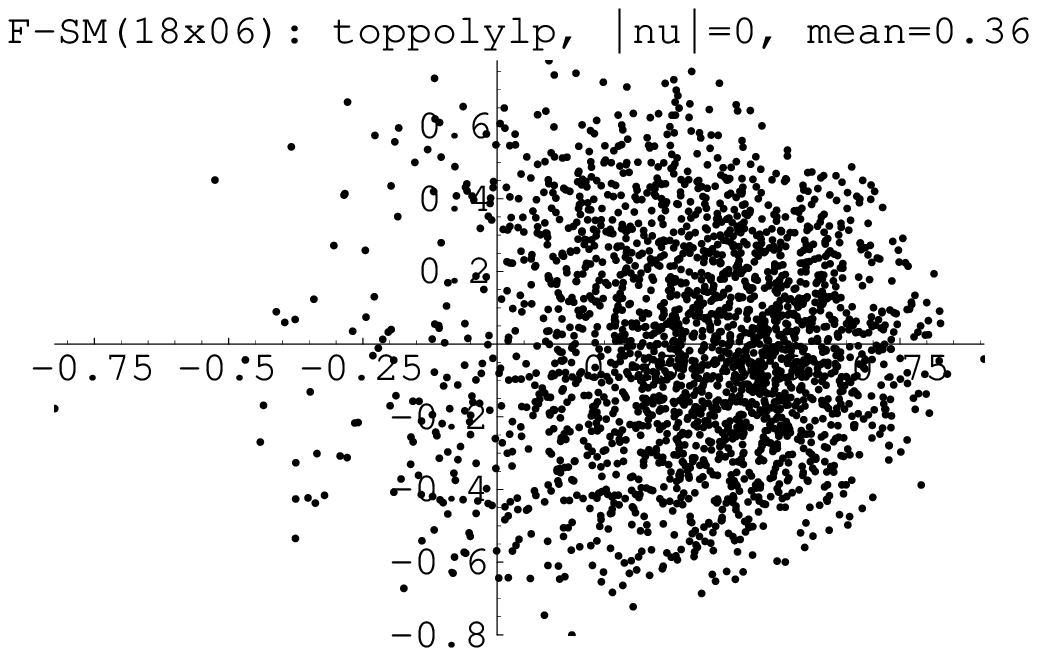,width=5.0cm}
\hspace*{1mm}
\epsfig{file=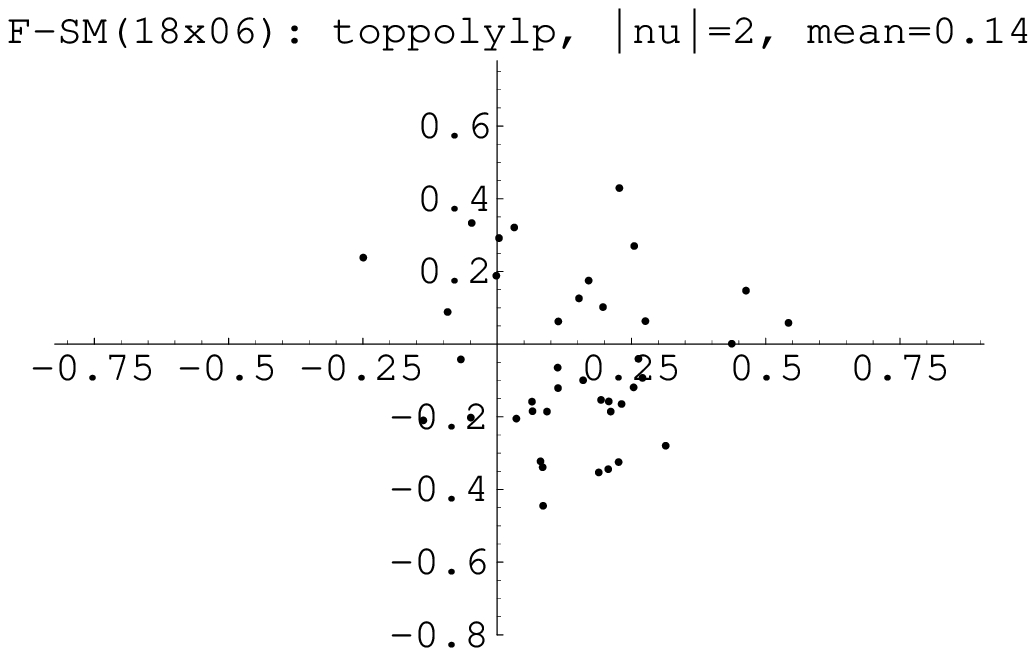,width=5.0cm}
\hspace*{1mm}
\epsfig{file=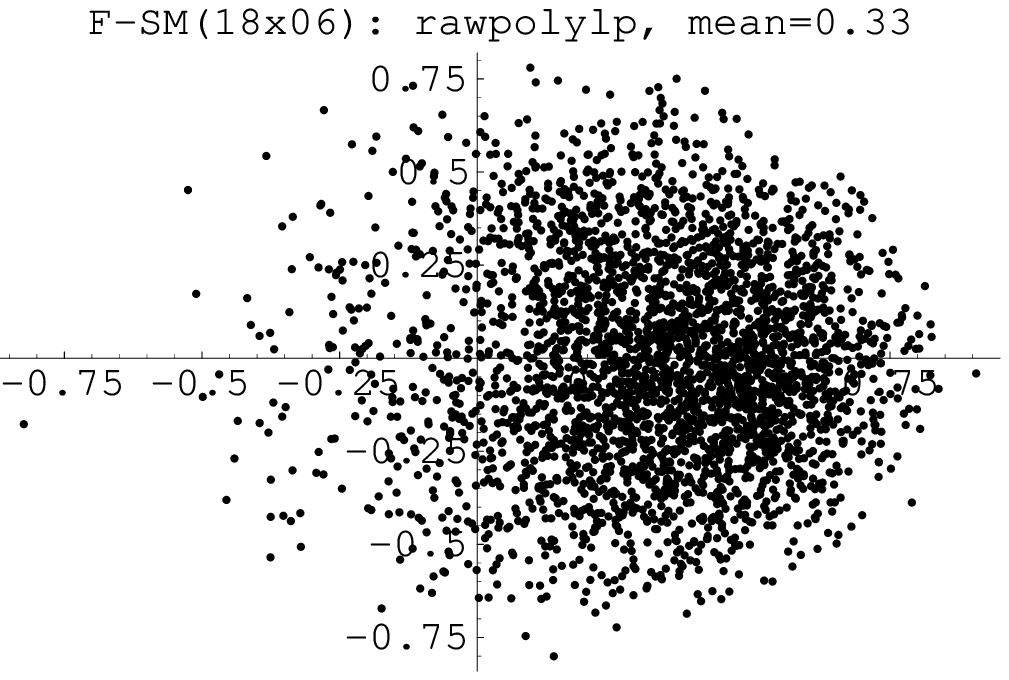,width=5.0cm}
\\
\epsfig{file=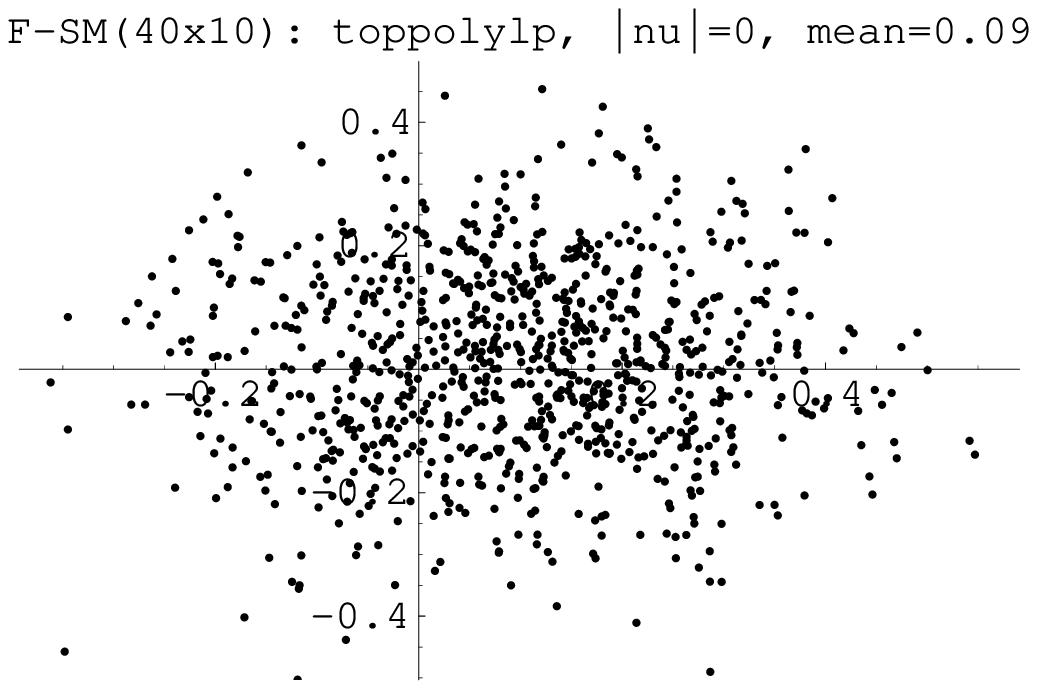,width=5.0cm}
\hspace*{1mm}
\epsfig{file=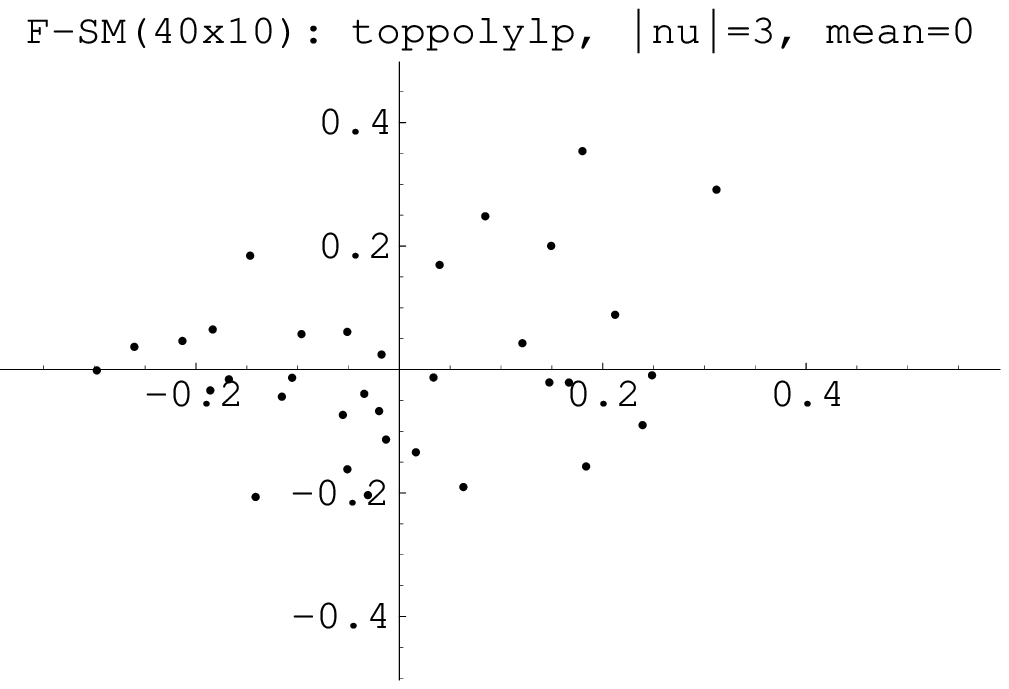,width=5.0cm}
\hspace*{1mm}
\epsfig{file=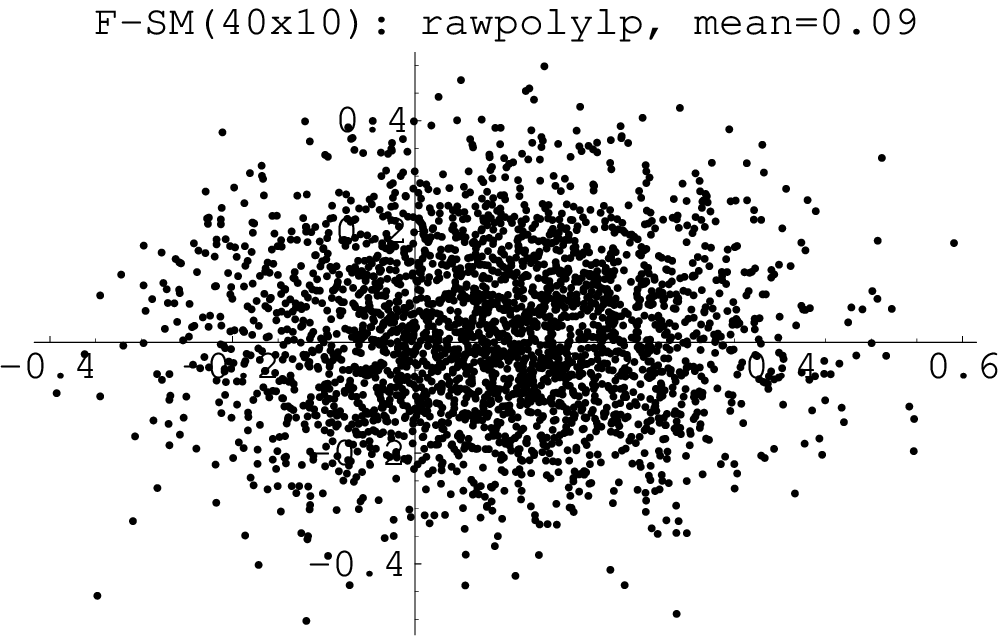,width=5.0cm}
\\
\vspace*{-12mm}
\caption{
Scatter plots of the sectoral expectation values $\<L\>_{\pm\nu}$ of the
Polyakov loop for the smallest and largest (reasonably populated) $|\nu|$ in
the small (8x4 lattice), intermediate (18x6) and large (40x10)
Leutwyler-Smilga regimes, respectively
(QED(2) data for $\beta\!=\!3.4, m\!=\!0.09, N_{\!f}\!=\!2$).}
\end{figure*}

\section{SECTORAL HEAVY QUARK FREE ENERGIES}

A question which has not been addressed so far is whether the
statements by Leutwyler and Smilga regarding the $\nu$-dependence of the
partition function (which we found well obeyed) would carry on to observables.
In other words: The question is whether in the small $x$ regime a typical%
\footnote{Here we shall exclude observables that obey a topological selection
rule, e.g.\ those that receive only contributions from the sectors with
$\nu\!=\!\pm1$, as is the case for the fermion condensate in the massless
limit.}
observable would be dominated by the contribution from the topologically
trivial sector and likewise, whether the same observable would prove
approximately independent of the topological charge of the background, if
the analysis is performed in the large $x$ regime.

A quantity which is easy to determine in lattice studies and which might help
elucidating the physical meaning of the LS-classification is the Polyakov loop,
i.e.\ the trace of a chain of link variables which winds once around the torus
in the euclidean timelike direction.
Its logarithm represents (up to a factor) the free energy of an external
(``heavy'') quark which is brought into the system without possibility to
influence it through back-reactions.
As in the previous section the analysis shall be done {\em sectorally\/},
i.e.\ on classes of configurations with a fixed value of $|\nu|$.
This means that the physical (unseparated) expectation value is understood as
a weighted mean
\beq
\<L\>=\sum_{|\nu|\ge0}\,p_{|\nu|}\,\<L\>_{|\nu|}
\label{PolyloopWeighted}
\eeq
of sectoral expectation values $\<L\>_{|\nu|}$, where the factor $p_{|\nu|}$
reflects the combined weight of the sectors $\pm\nu$ in the appropriate
histogram in Fig.~2.
On a practical level, the separation (\ref{PolyloopWeighted}) is facilita\-ted
by the fact that the ensemble is already {\em repre\-sentative\/} in the sense of
the full theory:
In order to compute the ensemble average [i.e.\ the practical version of the
l.h.s.\ of eqn.\ (\ref{PolyloopWeighted})], all one has to do is determine the
Polyakov loop on each configuration separately (where it is a complex number),
and then take the {\em arithmetic average\/} (which is supposed to be
approximately real).
The same holds true for the sectoral ensemble averages [i.e.\ the practical
versions of the $\<L\>_{|\nu|}$ showing up on the r.h.s.\ of
(\ref{PolyloopWeighted})].
Hence, in the case of the small $x$ simulation, instead of computing the sample
average from the 3197 indexed configurations directly (which gives
$L\!\simeq\!0.495$), one can also determine $\<L\>_0$ and $\<L\>_1$ from the
3052, 145 configurations with $|\nu|\!=\!0, 1$ (which gives 0.511 and 0.159)
and then combine these sectoral averages with the weights $p_0\!=\!3052/3197$,
$p_1\!=\!145/3197$.
In the run representing the intermediate LS-regime, the sectoral averages are
$\<L\>_0\!=\!0.361$, $\<L\>_1\!=\!0.228$, $\<L\>_2\!=\!0.142$, which together
with the weights $p_0\!=\!2343/3163$, $p_1\!=\!780/3163$, $p_2\!=\!40/3163$
reproduce the ensemble average $\<L\>\!=\!0.326$.
In the large $x$ simulation, the corresponding numbers are
$\<L\>_0\!=\!0.0947$, $\<L\>_1\!=\!0.0861$, $\<L\>_2\!=\!0.0777$,
$\<L\>_3\!=\!0.0003$ and $p_0\!=\!924/2504$, $p_1\!=\!1253/2504$,
$p_2\!=\!293/2504$, $p_3\!=\!34/2504$, from which one also gets the ensemble
average $\<L\>\!=\!0.0871$.

What one gains from this exercise is the insight that in the small $x$ regime
the physical expectation value $\<L\>$ is composed of sectoral averages
$\<L\>_0, \<L\>_1$ which prove very much inconsistent, i.e.\ in the symmetry
restoration regime a typical observable depends quite drastically on the
topological charge of the background.
In the intermediate LS-regime differences between neighboring topological
sectors are much smaller, but the system is still far from showing
overall consistency among all $\<L\>_{|\nu|}$.
In the large $x$ regime differences between neighboring topological
sectors happen to be marginal, but --~due to the highest topological sector
with $|\nu|\!=\!3$~-- data indicate that even in the large LS-regime some
limitations to the claimed \cite{LeutwylerSmilga} insensitivity on topology may
persist.


\section{SUMMARY AND CONCLUSION}

The net outcome of the investigation presented here is that the predictions
by Leutwyler and Smilga \cite{LeutwylerSmilga} regarding the overall
distribution of the topological charge $\nu$ in the regimes $x\!\ll\!1$
and $x\!\gg\!1$ are well reproduced.
Furthermore, the data ensure that the large $x$ regime (where SSB of the axial
flavour symmetry is pronounced) is special in several respects:
On a formal level the situation for $x\!\gg\!1$ is unique, as the functional
determinant results only in a (mild) overall suppression of higher topological
sectors w.r.t.\ lower ones, and the effective renormalization of $\beta$
is uniform, i.e.\ independent of $|\nu|$.\
On a more phenomenological level, the analysis of the sectoral dependence of a
typical observable has shown that the regime $x\!\gg\!1$ is unique, since here
physics proves (up to a limitation discussed above) independent of the
topological charge of the background.
Remarkab\-ly, our results illustrate and extend those of \cite{LeutwylerSmilga},
even though in our $x\!\simeq\!1$ and $x\!\gg\!1$ simulations the condition
$M_\pi L\!\ll\!1$ in the LS-analysis has been reversed into $M_\pi L\!\gg\!1$,
i.e.\ the pion would fit into the box.


\end{document}